\documentstyle[12pt]{article}
\begin{document} 

\def\beq{\begin{eqnarray}}
\def\eeq{\end{eqnarray}}

\noindent{\bf Quantum Gravity}

\noindent{Testing time for theories}
\vskip 0.75cm
\noindent{{\bf D. V. Ahluwalia}

\vskip 1.5cm

\noindent
{\bf Theories of quantum gravity attempt to 
combine quantum mechanics -- 
which reigns supreme at the atomic 
scale -- and the classical theory of general 
relativity, which governs planets and galaxies. 
The era of quantum gravity began about 
50 years ago with the pioneering
work of Chandrasekhar, when it was realized that 
quantum mechanics and gravitation combine 
in a fundamental  manner to form white 
dwarfs -- the Earth-sized stars of roughly a 
solar mass. That singular discovery had 
to wait  another 25 years for the  excitement to
begin afresh with the ideas put forward by 
Bekenstein, Hawking, Unruh, and Wheeler, 
among others. Yet, the present activity runs 
the danger of evolving into a mathematical 
science fiction in the absence of an experimental 
counterpart. For the last quarter of a 
century an important, though limited,
experimental quantum-gravity programme 
has existed. Giovanni Amelino-Camelia's 
paper on page 216 of  this issue\cite{gac} adds to this 
in a fundamental and significant manner  
by suggesting that the fuzziness of space-time 
is experimentally accessible.  Using existing 
data from a certain type of gravity wave detector
he is  able to set significant bounds 
on proposed quantum properties of space-time.
Instruments under construction 
promise to take us far beyond.}

If the universe is structured the way theorists
believe, the quantum mechanical unification
of the non-gravitational and gravitational interactions
must occur at a particular energy scale, called the
Planck scale -- just as the electromagnetic, strong,
and weak nuclear forces are predicted to unify at
a particular energy scale. But the actual scale at which theories
of quantum gravity can be explored has been poorly constrained.
The extreme smallness of both the Planck length ($\sim 10^{-35}$ m),
on the one side, and the ratio of the gravitational  
to the electrical forces acting between, 
say, two electrons, on the other side
($2.4 \times 10^{-43}$)
has led to the wide spread belief that 
the realm of quantum gravity is beyond 
terrestrial experiments. But, the following elementary 
facts can help dispel this view. For 
terrestrial experiments, a highly  relevant 
dimensionless quantity is derived from  the  
Newtonian gravitational potential 
$\phi_{grav}$ and 
the speed of light $c$. The constructed dimensionless 
object is:
\[\Phi_{grav}\equiv \left[\phi_{grav}/ c^2\right]_{Earth} = -\, 6.95 
\times 10^{-10}\] 
This is about 33 orders of magnitude larger 
than the ratio of the gravitational to the electrical 
force. Moreover, for quantum-gravity 
experiments it is not only the force that is of 
relevance, but also the  phases of the wave functions
of quantum mechanical  states.
Although in the non-relativistic  weak-field 
limit, the force depends on the gradient of 
$\phi_{grav}$, the phase depends directly on $\phi_{grav}$. In 
many situations, phase measurements give 
an enormous advantage and are also capable 
of highlighting the differences between the 
quantum and classical realms of gravity.
Therefore, for a physical state appearing as a 
linear superposition of different mass (or, energy)
eigenstates, non-trivial, gravitationally-induced
phases can come into play, and in principle are measurable.

Gravity waves are an important prediction
of Einstein's theory of general relativity
and though not yet detected directly, may 
provide information about the very early universe.
Although the  Planck length is indeed extremely small, 
modern gravity-wave interferometers are designed to detect
minute displacements in the positions of some test masses
(relative to a beam-splitter). There 
are additional consideration that may 
counterbalance the smallness of the Planck 
length. In certain quantum-gravity theories
the quantity $c^2 f^{-2} \lambda_{Planck}$ ($f=\mbox{frequency}$) 
characterizes the fuzziness in the distance 
between the test masses used in the experiment, 
whereas the quantity $c\, S(f)^2$  (where 
$S(f)$ is the amplitude spectral density of 
the  gravity-wave interferometer) characterizes 
the level of sensitivity to such fuzziness that 
modern gravity-wave detectors can achieve.
For frequencies of a few hundred hertz the 
two indicated quantities are comparable.

An experiment that exploited some of 
these observations, and which still remains 
relatively unknown in the quantum gravity 
community, is the 1975 classic experiment of 
Colella, Overhauser, and Werner (COW). It 
simultaneously explored the quantum and 
gravitational realms  using neutron interferometry \cite{COW}. 
To the accuracy in phase shifts of 
about 1\% available at that time, the COW 
experiment established the non-relativistic
weak-field limit of any viable theory of quantum 
gravity. For a single mass eigenstate, the 
amplitudes associated with each of the paths 
in a COW interferometer (to be distinguished from interferometers 
designed to detect gravity waves) 
picks up a different 
gravitationally induced phase, and so result in an 
observable change in the interference 
pattern as one path is rotated with respect to 
the other. In  this table-top  experiment the 
rotation is carried out in  such a manner that  
each path experiences a different gravitational 
potential. 

However, as the experimental accuracy 
has improved, the latest (1997) experiments 
show that a statistically  significant  discrepancy 
has begun to appear between the theoretical 
prediction of the gravity-induced phase shift
and that which is experimentally observed \cite{dis}.
On the other hand, when one 
studies effects of gravity in certain atomic 
systems, where the experimental measurements 
are about five orders of magnitude 
superior, no statistically significant discrepancies 
are observed \cite{atomic}. 
In the latest such experiments 
gravitationally induced quantum 
effects of local tides at Stanford have been 
observed.

Is there a slight difference in the manner 
in which neutrons and electrons interact 
with gravity, and can this point a way 
towards an appropriate expression of quantum 
gravity? In recent years, while the new 
generation of COW  experiments was reaching 
these higher levels of insight, three more 
quantum-gravity experiments have been 
proposed. These experiments are based on 
the possibility that quantum gravity might
affect the nature of fundamental symmetries,
or that the theory of general relativity itself
may not provide a complete description
of gravitation. One experiment is based 
on tests of CPT (the combined symmetries of
particle-antiparticle, parity, and time reversal)
invariance using the very sensitive 
neutral-kaon system \cite{neutralk}. The second 
concerns the possibility that quantum gravity might 
deform Lorentz symmetries in a way that 
would  alter the propagation of the $\gamma$rays we 
collect from astrophysical sources \cite{grb}. The third 
test explores  the incompleteness of the 
theory of general relativity itself, and proposes 
to measure quantum mechanically any constant gravitational potential 
in which we may be embedded \cite{dva}.

Here we  come to the Amelino-Camelia paper \cite{gac},
 which opens up yet another realm of
quantum gravity to experiments. The experiments 
he proposes would probe what is perhaps the central
element of quantum gravity, 
namely, the concept of space-time.
He notes that almost all existing approaches to quantum 
gravity expect a modification of the classical picture 
of space-time by introducing some sort of fuzziness. 
With that we all agree, 
but there are various proposals as to what 
concrete form this fuzziness takes. 
Amelino-Camelia convincingly argues that gravity-wave 
interferometers have precisely the right 
ability to probe the space-time fuzziness. 
He sets highly significant limits (in one 
case ultra-Planckian) on the length scales characterizing 
two fuzziness situations. Amelino-Camelia's 
paper exploits the differences in 
the fuzziness of the distance between two test 
masses used in the experiment, as predicted by 
different theories of quantum gravity.
It turns out that these differences are large
enough to be detected in the experiments being proposed.
So, whereas the smallness of the Planck length and the ratio
of gravitational to electrical forces, 
does play its
own essential role in nature, it does not make quantum gravity a 
science where humans cannot venture  to probe her
secrets.

\medskip
\noindent{\it D. V. Ahluwalia is at the 
Escuela de Fisica, Univ. Aut. de Zacatecas,
Apartado Postal C-580, Zacatecas, ZAC 98068, MEXICO}

\noindent{\it e-mails: ahluwalia@phases.reduaz.mx, av@p25hp.lanl.gov}

\end{document}